# A Performance Analysis Tool for Nokia Mobile Phone Software

Edu Metz, Raimondas Lencevicius
*Nokia Research Center*
*5 Wayside Road, Burlington, MA 01803, USA*
*Edu.Metz@nokia.com Raimondas.Lencevicius@nokia.com*

**ABSTRACT**

Performance problems are often observed in embedded software systems. The reasons for poor performance are frequently not obvious. Bottlenecks can occur in any of the software components along the execution path. Therefore it is important to instrument and monitor the different components contributing to the runtime behavior of an embedded software system. Performance analysis tools can help locate performance bottlenecks in embedded software systems by monitoring the software's execution and producing easily understandable performance data. We maintain and further develop a tool for analyzing the performance of Nokia mobile phone software. The user can select among four performance analysis reports to be generated: average processor load, processor utilization, task execution time statistics, and task execution timeline. Each of these reports provides important information about where execution time is being spent. The demo will show how the tool helps to identify performance bottlenecks in Nokia mobile phone software and better understand areas of poor performance.

## 1. Introduction

Performance is one of the most important non-functional requirements of embedded software systems. As the complexity of software in mobile phones grows, software performance debugging and analysis becomes an important issue. Software performance debugging and analysis is about looking at the software execution to pinpoint where bottlenecks occur. Bottlenecks are places in the execution path that require an inordinate amount of time relative to the execution of an entire operation. Trace based performance analysis tools can help to discover performance bottlenecks in embedded software systems by monitoring the software's execution and produce easily understandable performance data. We maintain and further develop a tool for analyzing the performance of Nokia mobile phone software. The tool provides an easy way for users to find out where time is being spent and to understand and correct performance problems.

The remainder of this paper describes the approach taken to analyze the performance of Nokia mobile phone software, the performance analysis tool, and then demonstrates the capabilities of the tool by providing and analyzing performance report examples.

## 2. Performance Analysis

The approach taken to analyze the performance of Nokia mobile phone software is to collect performance data during software execution through software tracing and then provide post-mortem analysis and display of performance information. Permanent trace hooks inserted in the operating system kernel are used to record information about performance related events in the software execution. The operating system kernel reports these events through a dedicated tracing interface. The user does not have to change the mobile phone software, just to activate the operating system kernel hooks at compile time. There exists an interface that allows the user to specify which type of trace hooks are of interest, and to activate only these hooks. Once the instrumented software build is loaded onto the target hardware, test cases are executed, produced traces are collected and stored in trace files. A simplified sample trace file is depicted in Figure 1. The trace file consists of event traces with the following format: *<timestamp> event type: event data*. After generating the trace file, the collected traces are processed by the performance analysis tool. The tool outputs different performance analysis reports in plain-text tabular format. Excel enhanced with Visual Basic macros is used as a front-end tool for graphical visualization of the report data. Figure 2 shows the data flow diagram of the performance analysis approach.

---

```
<0000h 00m 01s 290 602> Task schedule: old 5 new 3
<0000h 00m 01s 290 678> Task schedule: old 3 new 1
<0000h 00m 01s 290 764> Task schedule: old 1 new 4
<0000h 00m 01s 290 838> IRQ begin: 16
<0000h 00m 01s 290 861> IRQ end: 16
<0000h 00m 01s 290 922> Task schedule: old 4 new 2
<0000h 00m 01s 291 015> Task schedule: old 2 new 5
<0000h 00m 01s 291 091> IRQ begin: 23
<0000h 00m 01s 291 124> IRQ end: 23
<0000h 00m 01s 291 230> Task schedule: old 5 new 3
```

Figure 1: Trace file sample

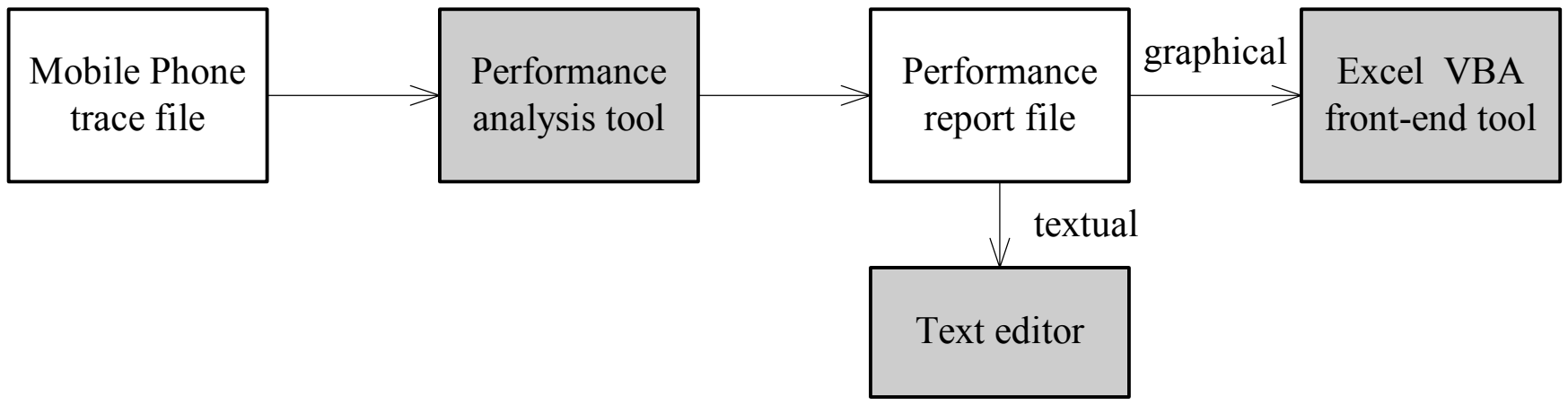

Figure 2: Performance analysis approach overview

Currently our trace tool collects only task-level scheduling data. Function-call level traces may be interesting and important too, but there is an inherent tradeoff concerning the amount of trace data collected, the types of analysis that can be performed on them, and the overhead of the data collection [Rei01]. Trace instrumentation changes the performance of a software system. Not only does event tracing take some time, adding traces may change the behavior of the software system. In addition, in a real-time software system, instrumentation could possibly result in violation of real-time constraints and timing requirements. Trace instrumentation reduces the validity of the performance analysis; therefore, there is a need to minimize the effect of the tracing on the underlying system's performance. These issues are explored in more detail in our paper [Met03].

## 3. Performance Analysis Tool

Performance analysis tools are a key component in the development of Nokia mobile phone software. The success of a performance analysis tool depends on the tool's ability to process and analyze performance data and relate this information in terms meaningful to the user. It is relatively easy for a developer to improve the execution time of an individual software component, but when improving overall system performance a lot of effort can be wasted in the wrong place – tuning components that have only a small impact on overall system performance. Off-the-shelf performance analysis tools are not suited for analyzing the performance of Nokia mobile phone software. There are many reasons for this. First, the trace instrumentation in Nokia mobile phone software is proprietary and off-the-shelf performance analysis tools do not have capabilities to process and analyze these proprietary traces. Second, standard profiling tools, such as Gprof [Gra82] and Rational Test RealTime [Rat03], provide useful performance data, but they only attribute runtime cost to components at the function level.

Therefore, an in-house tool for analyzing the performance of Nokia mobile phone software was developed. The tool is a Perl based post-mortem analysis tool capable of analyzing the execution of tasks and interrupt handlers. The tool works with a trace file containing operating system scheduling data and processes the data to extract relevant performance information. For example, if the sample file depicted in Figure 1 is given as input file, the tool calculates that IRQ handler 16 executes for 43 µs and task 4 executes for 135 µs. Based on user-provided command line options, the tool generates different performance analysis reports: average processor load, processor utilization, task statistics, and task execution timeline. Each performance analysis report is output in a plain-text tabular format and can be presented in two ways: textual and graphical representation. An Excel Visual Basic front-end tool is used to make a graphical representation of the report. This front-end tool allows us to take advantage of Excel's graph capabilities and its interface known to almost all users.

Each of the four performance analysis reports: average processor load, processor utilization, task statistics, and task execution timeline, individually provides important information about where the execution time is being spent. Average processor load reports the average utilization of each task executed during the use scenario and more importantly the average amount of processor idle time. Processor utilization reports what task is executing at what time. It reports the relative execution time of a task at each timeslot interval over a period of time. This information could be helpful when debugging timing errors and it helps to make scheduling decisions. Analyzing the task's processor utilization is a necessary step toward fully understanding the timing of an embedded software system [Ste01]. Task statistics reports detailed execution time statistics for each task executed during the use scenario. This report provides the user with important information such as the percentage of the total amount of time spent in each task, minimum, worst case, and average execution time of each task, and exponential and uniform distribution for task execution times and task periods. This information helps to analyze the temporal properties of real-time tasks in mobile phone software. Analysis of tasks' execution time distributions is the basis for most scheduling analysis [Ran03][Ste01]. Task execution timeline reports task states over time. This provides visibility into the execution order of tasks.

The intent of our performance analysis tool is to provide an easy way for the user to identify performance issues in Nokia mobile phone software and understand and correct areas of poor performance. The performance issues of concern addressed by the tool include: timing and real-time scheduling problems, overhead of wireless communication protocols, energy consumption, and antipatterns with negative software performance consequences, such as unnecessary and extensive processing [Smi02]. The tool does not automate the search for performance issues – it only assists in the search. It is the responsibility of the user to use the displayed information to analyze the performance of Nokia mobile phone software. This can be a non-trivial task, because the user needs to identify the factor that is affecting the performance. Automated performance analysis of mobile phone software would require a significant investment.

To understand the capabilities of the tool, let us look at a few different performance analysis reports of Nokia mobile phone software produced by the tool. The scenario discussed, its trace, and task execution parameters in the trace are simplified for presentation purposes. Figure 3 illustrates the graphical representation of a processor utilization report. The relative execution time of each task at a given time is highlighted through a coloring scheme. Task 0 represents the idle time of the processor. The task statistics report provides detailed execution time statistics for each task executed during the use scenario. Distribution of a task's execution time is one type of information contained in this report. Figure 4 shows the graphical representation of exponential distribution of execution time and period for task 1. Another type of information provided by the task statistics report includes relative amount of time spent in each task as well as minimum, worst case, and average execution time of each task. Figure 5 lists some of the execution time information for task 1.

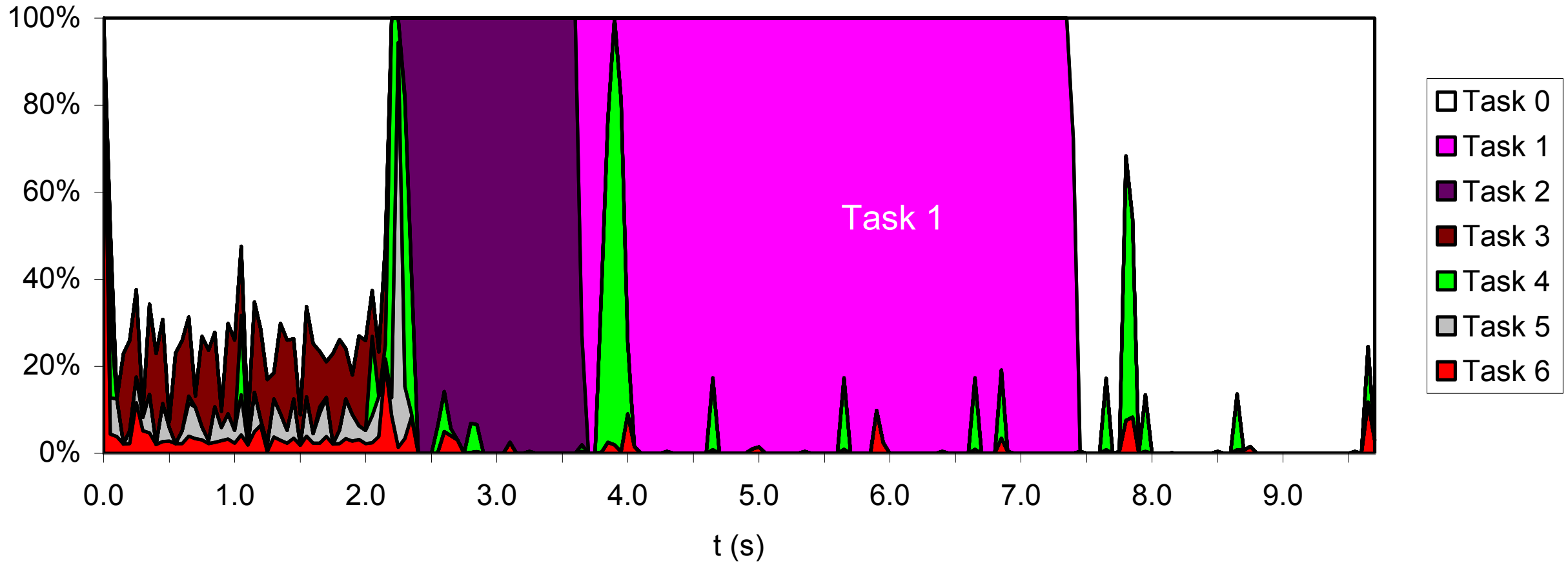

Figure 3: Processor utilization

Analysis of the processor utilization starts with a view from the entire use scenario (Figure 3). When necessary, the user can view the utilization for a specific time slice by using the zoom capability of the performance analysis tool. In Figure 3 the execution time of task 1 stands out from the rest. It is apparent that task 1 uses most of the processor within the 3.5 to 7.5 second timeframe. Figure 4 shows that most of the execution times of task 1 are almost as long as the corresponding period. The worst-case execution time of task 1 is calculated to be 526 ms (Figure 5). So the obvious question that arises: is the execution behavior of task 1 correct? At first sight the execution behavior of task 1 seems to be incorrect. In mobile phone software long running tasks should be avoided because they could prevent other (real-time) tasks from executing. It turns out that task 1 is a background task running at a low priority and it only executes when the processor is idle. Therefore, the execution behavior of task 1 is correct.

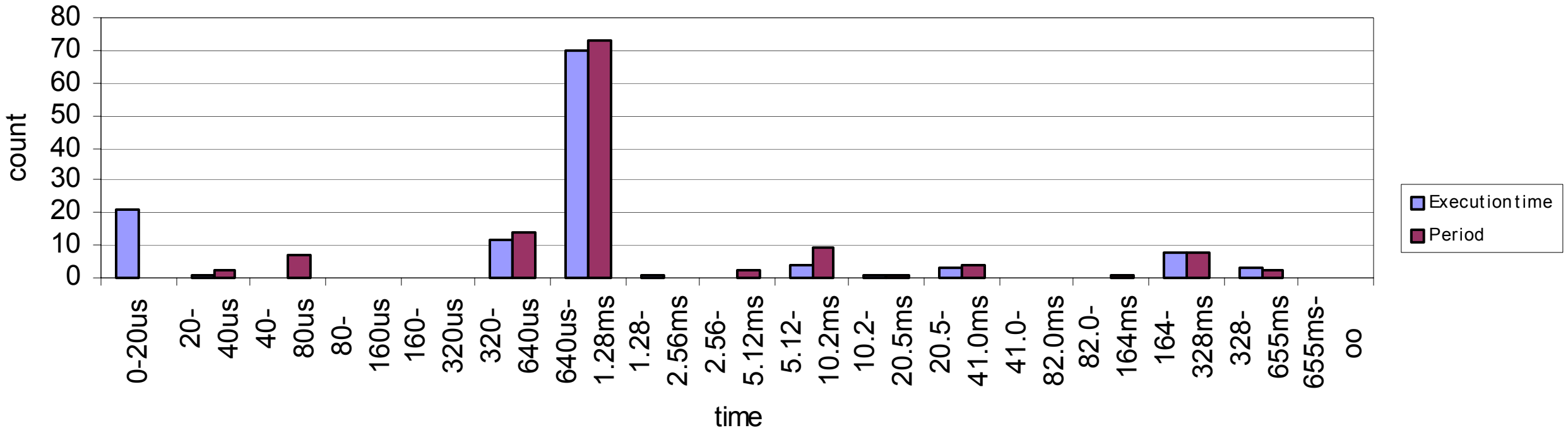

Figure 4: Time distribution histogram for task 1

| | |
|---|---|
| Average utilization | 36.85% |
| Worst case execution time | 526 ms |
| Minimum execution time | 6 usec |
| Average execution time | 29 ms |

Figure 5: Execution time information for task 1

## 4. Summary

We have described a post-mortem performance analysis tool for Nokia mobile phone software. The tool is able to analyze and report performance information in four different ways: average processor load, processor utilization, task execution time statistics, and task execution timeline. Each of these reports provides important information about where execution time is being spent. The tool helps users to identify software performance bottlenecks and better understand the performance characteristics of a software system.

The demo will demonstrate the capabilities of our performance analysis tool by providing and analyzing performance reports of Nokia mobile phone software.